\documentclass{article}

\usepackage{caption}
\usepackage{subfigure}
\usepackage{PRIMEarxiv}
\usepackage{float}
\usepackage[utf8]{inputenc} 
\usepackage[T1]{fontenc}    
\usepackage{hyperref}       
\usepackage{url}      
\usepackage{booktabs}       
\usepackage{amsfonts}       
\usepackage{nicefrac}       
\usepackage{microtype}      
\usepackage{lipsum}
\usepackage{fancyhdr}       
\usepackage{graphicx}       
\graphicspath{{media/}}     
\usepackage{amsmath}
\usepackage{multicol}

\title{Extract fundamental frequency based on CNN combined with PYIN

}

\author{
  Ruowei Xing \\
  Xi’an Jiaotong-liverpool University \\
  \texttt{ruowei.xing19@student.xjtlu.edu.cn} \\
   \And
  Shengchen Li \\
  Xi’an Jiaotong-liverpool University \\
  \texttt{Shengchen.Li@xjtlu.edu.cn} \\
}

\begin{document}
\maketitle

\begin{abstract}
This paper refers to the extraction of multiple fundamental frequencies (multiple F0) based on PYIN, an algorithm for extracting the fundamental frequency (F0) of monophonic music, and a trained convolutional neural networks (CNN) model, where a pitch salience function of the input signal is produced to estimate the multiple F0. The implementation of these two algorithms and their corresponding advantages and disadvantages are discussed in this article. Analysing the different performance of these two methods, PYIN is applied to supplement the F0 extracted from the trained CNN model to combine the advantages of these two algorithms. For evaluation, four pieces played by two violins are used, and the performance of the models are evaluated accoring to the flatness of the F0 curve extracted. The result shows the combined model outperforms the original algorithms when extracting F0 from monophonic music and polyphonic music.
\end{abstract}

\keywords{F0 extraction\and PYIN\and CNN}

\begin{multicols}{2} 
\section{Introduction}
Automatic extraction of F0 and the harmonic frequencies has been one of the persistent challenges of music signal processing. The expensive cost of manually marking the F0 and lack of database that has complete F0 annotations lead to many inconveniences in music acoustics research such as timbre classification.\cite{marozeau2003dependency} \par
So far, many researches try to find a method to get the correct F0 value automatically. PYIN and YIN are two sophisticated approaches in extracting F0 from the monophonic music\cite{mauch2014pyin,de2002yin}, but they fail to extract the fundamental frequency of even a single tone from a polyphonic music. The conventional Multi- F0 tracking methods focus on heuristics, such as creating a F0 salience spectrum based on summing the amplitudes of harmonic partials \cite{klapuri2006multiple}, optimizing the maximum- likelihood according the harmonics and spectral peaks/non-peaks\cite{duan2010multiple}, and some other pitch salience function based approaches\cite{ryynanen2008automatic, salamon2012melody}. \par
Other approaches on tracking Multi-F0 are data-driven, based on train data labeled with manual F0 annotations. A recent example is a system called DeepSalience proposed by Bittner et al\cite{bittner2017deep}: A CNN model trained on harmonic constant-Q transform (HCQT), a 3-dimensional array indexed by harmonic index, frequency and time, to get the signal's pitch salience representation. However, although these Multi-F0 tracking methods extract the F0 successfully, their accuracy and precision are far lower than the manual annotations. \par
In order to improve the accuracy of the Multi-F0 tracking approach, the advantages of the existing tracking methods are combined. In this paper, we focus on supplementing and improving the Multi-F0 extracted from a CNN model proposed by Cuesta et al.\cite{cuesta2020multiple}, which outperforms the DeepSalience model\cite{bittner2017deep}, by using the F0 extracted from the PYIN.

\section{Method}
In the F0 tracking of the monophonic part of the music, PYIN is much better than other Multi-F0 tracking methods because it only focuses on detecting one F0 value. In contrast, the Multi-F0 extraction approaches perform better in the polyphonic part of the music. Hence, we choose a well-performed CNN system combined with PYIN to get a more accurate F0. 
\label{sec:headings}

\subsection{Database}
In the research, we use 4 pieces played by violin to evaluate the result of the extraction method. Of the four melodies, the same person played two different pieces on two violins. Most of one melody is monophonic, while another one contains many polyphonic parts. 

\subsection{PYIN}
PYIN, an improved version of the convention YIN, is a sophisticated approach in extracting F0 from the monophonic music. YIN calculates the difference 
$$ d_t(\tau) = \sum_{j}^{W} (x_j - x_{j+\tau})^2  $$
between signals which will be small when the signal translation is approximately close to the fundamental period, and then uses the difference smaller than a predetermined fixed threshold s(usually s = 0.1 or s = 0.15) to estimate the F0 value. PYIN follows this method but remains the useful information discarded before comparing with the threshold in YIN. The frequency-probability is then calculated from the remained candidate based on the hidden Markov model.\cite{fine1998hierarchical} The detailed information of PYIN is shown in Figure1 and \cite{mauch2014pyin}.  
\begin{figure}[H]
\centering
\includegraphics[scale = 0.5]{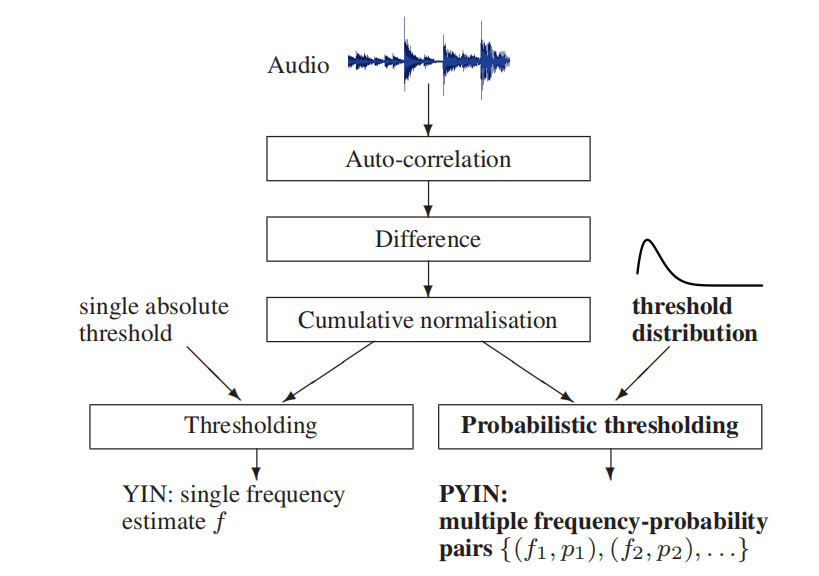}
\caption{ Comparison between the fundamental steps of the original YIN algorithm and the PYIN algorithm}
\end{figure}

\subsection{CNN Multi-F0 estimation}
The model is trained on the HCQT magnitude and the HCQT phase differentials, the detailed descriptions of them are illustrated in \cite{bittner2017deep,cuesta2020multiple}. The output of the model is a pitch salience spectrum, which corresponds to the representation of the trained HCQT. As all audio files are resampled to a sampling rate of 22050 Hz, and hop size is determined as 256 samples, we process the signal according to such standards to reduce deviation caused by different resolutions.\par
Trained by the polyphonic music, the model is more sensitive to the polyphonic part of the music, but it shows an equivalent performance to the simpler monophonic one, where PYIN shows an advantage. 

\subsection{Combined strategies}
To get the F0 that can be combined together, we set the signal frequency and time revolution of the two algorithms to the same. Here we use a sampling rate of 22050Hz, a window size of 1024 samples, a hop size of 256 samples. The F0 values extracted from the two algorithms respectively are assigned to the nearest frequency bin in the magnitude of short-time Fourier transform(STFT). The total process of the combination is illustrated in Figure 2.\par
As the CNN model is more sensitive to the polyphonic part and has a high accuracy regardless of some bias, the Multi-F0 extracted from it is assumed the correct result. Then F0 extracted from PYIN is compared with the F0 of the first tune ($M_1$) of the multiple F0. The combination is based on the difference between the two result:

\begin{equation}
\label{eq1}
	f_0(t)=\left\{
		\begin{aligned}
		&F_0(t)&, M_1[t] = NAN \\
		&M_1(t)&, M_1[t] \neq NAN
	\end{aligned}
	\right.
\end{equation}

\begin{equation}
\label{eq3}
	M_1(t)=\left\{
		\begin{aligned}
		&P[t] &&, \left|P[t] - M_1[t] \right| \leq 2bins\\
		&M_1[t] &&, else
	\end{aligned}
	\right.
\end{equation}

\begin{equation}
\label{eq2}
	F_0(t)=\left\{
		\begin{aligned}
		&P[t] &&, All[M_1[t+i] = NAN, i = -5,-4,\cdots, 4] \\
		&NAN &&, else
	\end{aligned}
	\right.
\end{equation}

The $f_0(t)$ here is just the F0 of the first tune of the final Multi-F0 at time t, and NAN means not a number(Fundamental frequency is not detected).$F_0[t]$ is the F0 value at time t extracted from PYIN, while $M_1[t]$ is the F0 of the first tune of the Multi-F0 extracted from the CNN model.

\begin{figure}[H]
\centering
\includegraphics[scale = 0.5]{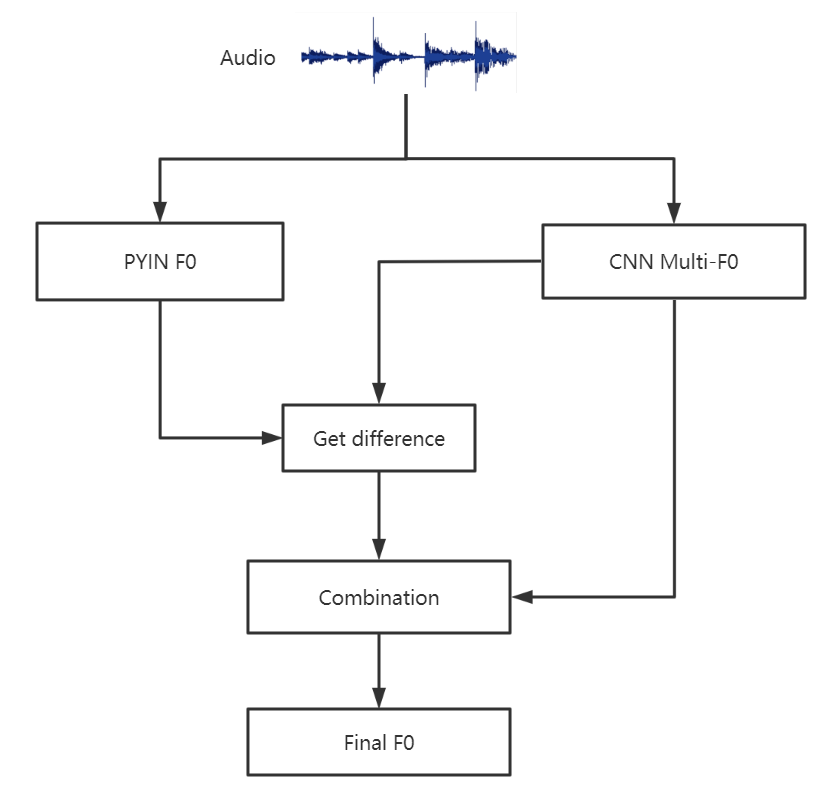}
\caption{Process of F0 combination}
\end{figure}

\section{Result}
\label{sec:others}
The F0 result extracted from the three algorithms are depicted in Figure 3. Although the combined result doesn't improve other F0 value, it supplements the first F0 in the Multi-F0 and replace the F0 of the monophonic part with a more precise F0. From the result, the curve of the combined F0 are smoother and more  complete than the other two algorithms.
\end{multicols}
\begin{figure}[htbp]
\centering 
 
\subfigure[Violin-a first melody: below] 
{
	\begin{minipage}{0.45\linewidth}
	\centering          
	\includegraphics[scale=0.4]{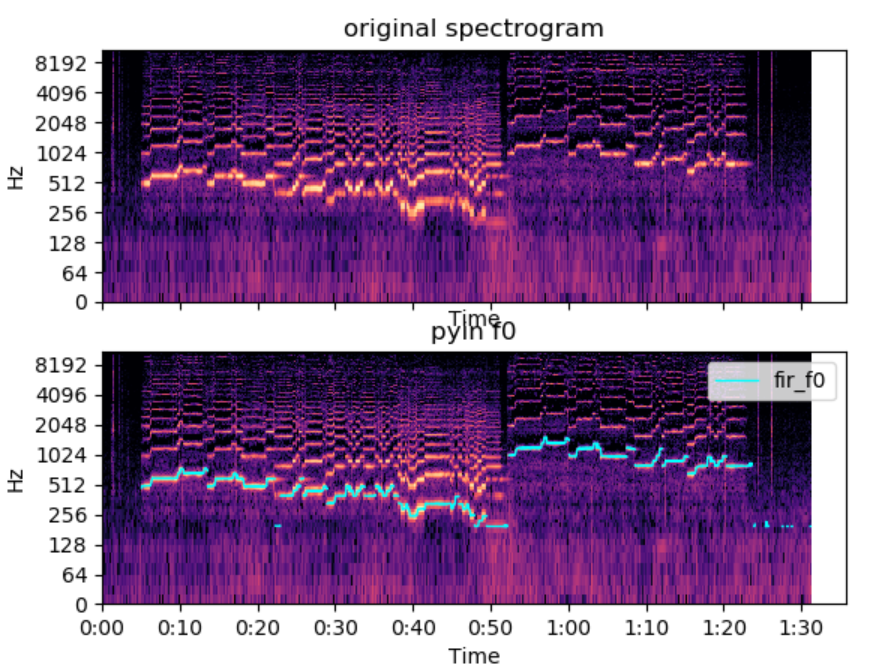}  
	\end{minipage}
}\quad
\subfigure[Violin-a first melody: below]
{
	\begin{minipage}{0.45\linewidth}
	\centering      
	\includegraphics[scale=0.4]{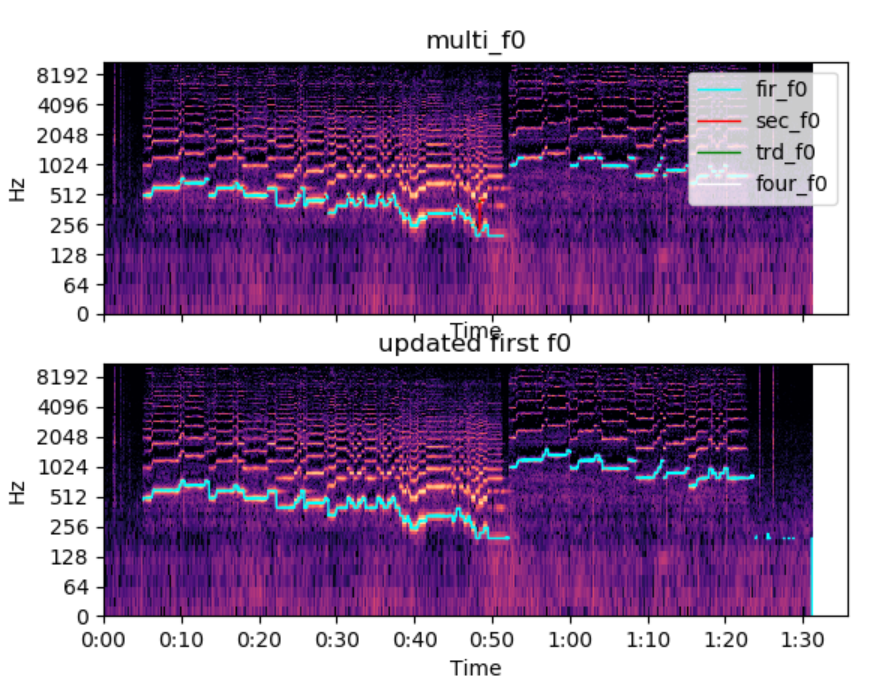}  
	\end{minipage}
}\quad
\subfigure[Violin-a second melody: original stft and F0 extracted by PYIN] 
{
	\begin{minipage}{0.45\linewidth}
	\centering         
	\includegraphics[scale=0.4]{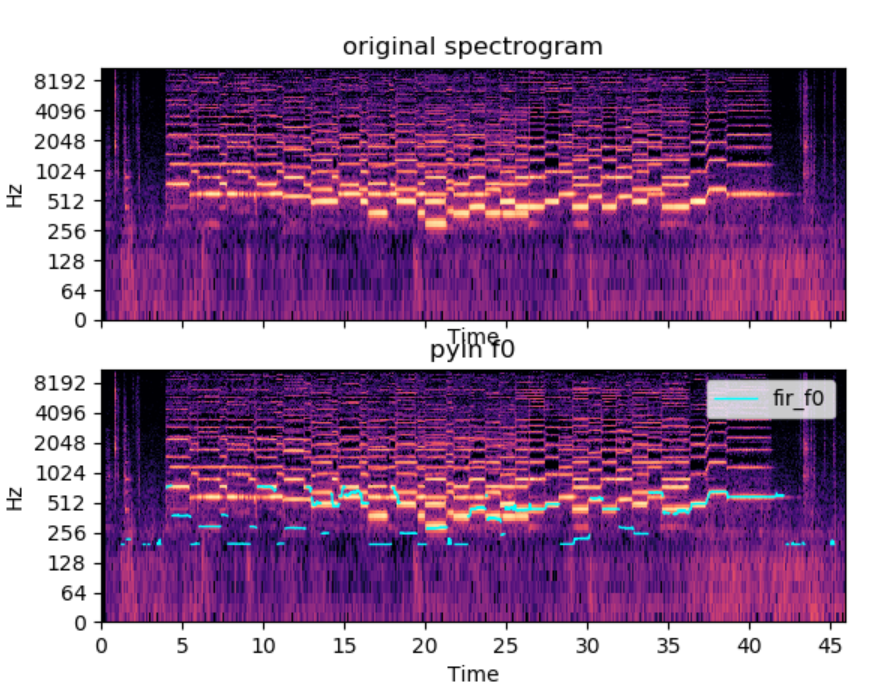}   
	\end{minipage}
}\quad
\subfigure[Violin-a second melody: Multi-F0 extracted by CNN model and the updated first f0. The other F0 except the first F0 is same in the two figure, so they are not depicted in the figure below] 
{
	\begin{minipage}{0.45\linewidth}
	\centering      
	\includegraphics[scale=0.4]{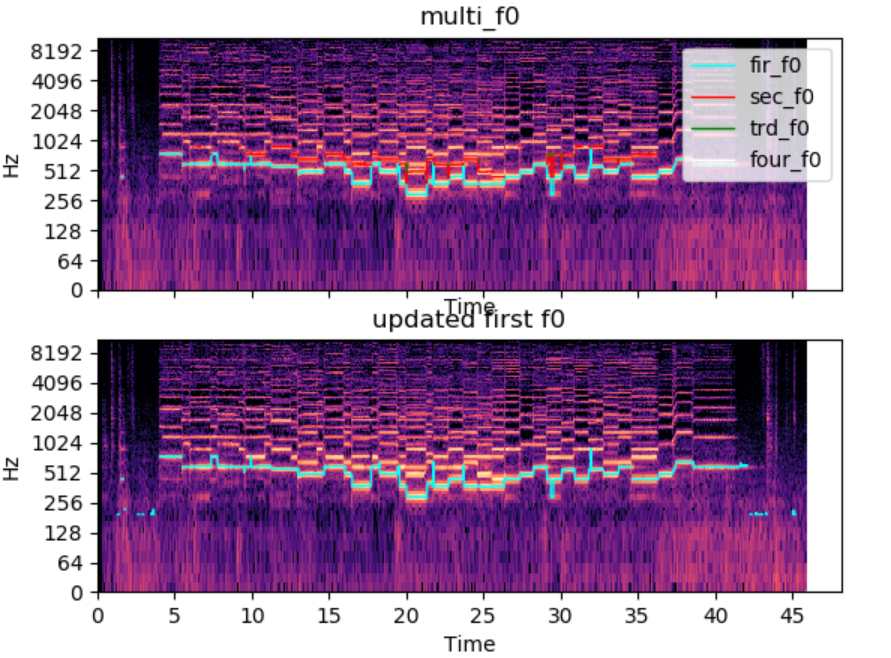}   
	\end{minipage}
}
 
\caption{Result of violin-a melody. Violin-b is similar to the violin-a}
\label{fig:1}  
\end{figure}

\section{Conclusion}
In this paper, we proposed a combined Multi-F0 extraction method, using the F0 extracted by PYIN and a CNN model. We use four pieces played by violins which contains both monophonic and polyphonic parts to evaluate the combined model, and get a smoother F0 curve with this new approach. 
Although this research just provides a simple process to combine the F0 extraction method, it tries to provide a way for the joint F0 extraction and get a better result successfully. In the future, modifying weight of the HMM in the PYIN according to the neural network output such as the output 0f the CNN model will be considered. Further steps include improving the performance of the neural network in Multi-F0 tracking and searching for some conventional F0 tracking method that can benefit from some rough results(output from neural network) before the tracking process.

\bibliographystyle{unsrt}  
\bibliography{references}

\end{document}